# There Are Two Distinct Photon Gases Present Inside Every Solar Cell

Eli Yablonovitch & Zunaid Omair

UC Berkeley Electrical Engineering & Computer Sciences Dept.

Berkeley, California, USA 94720-1770

## 1. Abstract:

It has gradually been recognized that incoming sunlight can be trapped within a high refractive index semiconductor, $n≈3.5$, owing to the narrow 16° escape cone. The solar light inside a semiconductor is $4n^2$ times brighter than incident sunlight. This is called light trapping and has increased the theoretical and practical efficiency of solar panels. But there is a second photon gas of equal importance that has been overlooked. Inside every forward-biased solar cell there is a gas of infrared luminescence photons, also trapped by total internal reflection. We introduce the idea of *super-equilibrium*, when the luminescence photon gas freely exchanges energy with the two quasi-Fermi levels.

Nonetheless, the loss of a single photon from either gas is equivalent to the loss of a precious minority carrier. Therefore optical modeling & design becomes equally important as electron-hole modeling in high efficiency solar cells. It becomes possible to approach the idealistic Shockley-Queisser limit, by proper material selection and design of the solar cell optics.

## 2. The Two Distinct Photon Gases:

Inside a solar cell, the incoming solar photons tend to be trapped by total internal reflection. Weakly absorbed rays of sunlight experience multiple internal reflections, forming a photon gas with $4n^2$ absorption enhancement [1], increasing both current and voltage.

At the same time, there is second photon gas inside every solar cell. The presence of a non-zero voltage is accompanied by an infrared luminescence photon gas, significantly brighter than Planck's black-body radiation.

These two photon gases are very important for solar cell operation. If a single photon is lost from either photon gas, it is equivalent to the loss of a precious minority carrier. Thus the photon gases need to be included in solar cell modelling, and must be treated on an equal footing with the electron gas, and with the hole gas, that are also present in every solar cell.

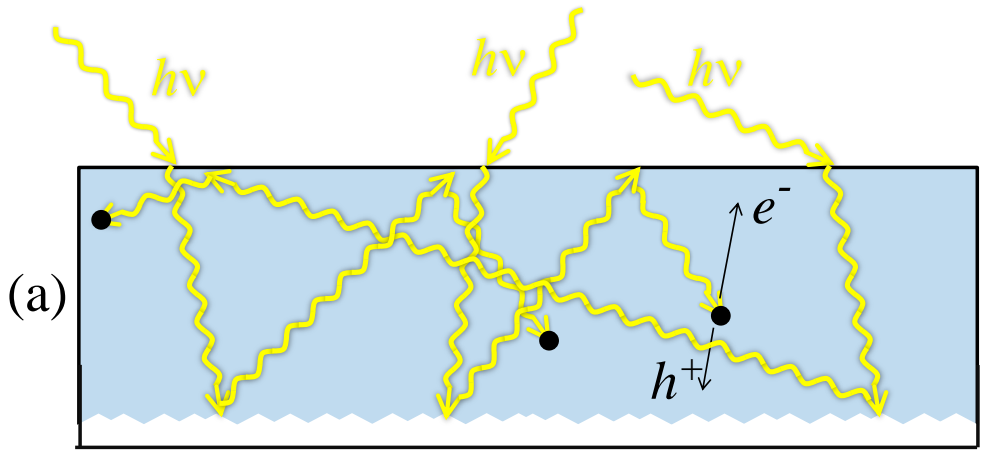

Figure 1(a): The photon gas formed by weakly absorbed solar photons, reflected from the random scattering rear surface. The internal brightness is $4n^2$ >than the incoming sunlight.

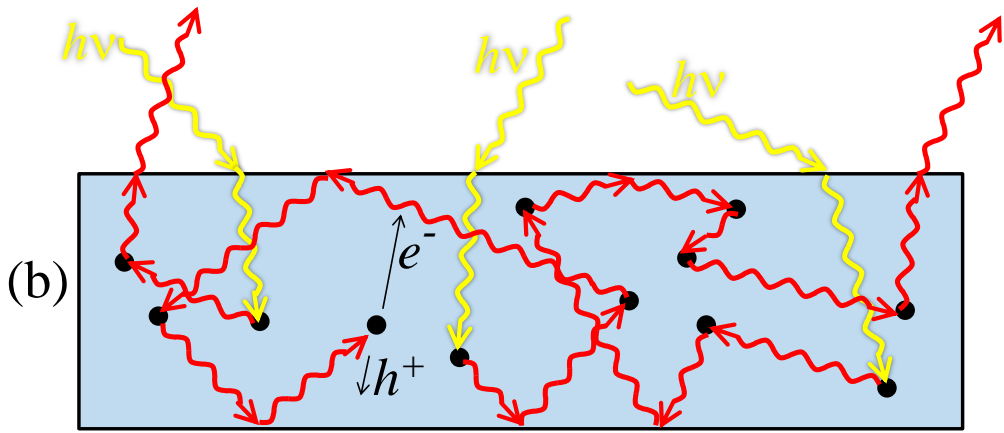

Figure 1(b): The other internal photon gas (in red) formed by luminescent infrared photons.

## 3. The Generalized Planck Theorem:

In thermodynamic equilibrium, there is only one Fermi Level (or Chemical Potential) for electrons in a system. Semiconductors are unusual in that they separately sustain both, an electron gas and a hole gas. In the best materials, the electrons and holes recombine rarely, and barely interact. The electrons and holes fail to mutually equilibrate, and each has a separate Fermi Level. Since they are out of global equilibrium, the separate Fermi Levels are called quasi-Fermi levels, $E_{Fn}$ and $E_{Fp}$ respectively, where $E_{Fn}-E_{Fp}=\mu=qV$. The quasi-Fermi level separation $\mu=qV$ represents the internal free energy buildup by sunlight.

In a solar cell there is a further type of equilibrium: The internal luminescent photon gas exchanges energy with the electrons and holes, and can establish a form of equilibrium that includes photons, electrons and holes. We suggest the name "super-equilibrium" when the infrared photon gas exchanges energy between the two quasi-Fermi Levels. The photon gas Brightness, B, then deviates from the Planck black-body formula:

$$\mathcal{B}(\nu,\mu,T) = \frac{8\pi n^2 \nu^2}{c^2} \frac{1}{e^{\frac{h\nu-\mu}{kT}}-1} \quad \ldots\ldots\ldots\ldots\ldots\ldots\ldots\ldots\ldots(1)$$

where brightness $\mathcal{B}$ is photon number/area, per unit bandwidth, per unit time, per $4\pi$ steradians, $n$ is the refractive index, and $kT$ the ambient thermal energy. This is sometimes called the "Generalized Planck formula". It was first derived by Ross [2] and further elaborated by C.H. Henry [3], and in ref. [4]. If there is no excess carrier concentration, and only a single quasi-Fermi level, there is global equilibrium, $\mu=qV=0$, and the ordinary Planck formula would apply. But if $\mu\neq0$, then the "super-equilibrium" would apply, as represented by the "Generalized Planck formula", eq'n. (1).

It is odd that a photon spectral distribution, eq. (1), would contain a Chemical Potential, $\mu$. Since photon number is not conserved, they can be freely created and destroyed, it is usually understood that creation of photons costs zero Free Energy, $\mu=0$. But that only applies to perfect thermal equilibrium. In our case, the photons are closely coupled with, and exchange energy with, the carriers of an excited semiconductor. The photon distribution, eq'n. (1), then contains the Chemical Potential of the semiconductor carriers, the quasi-Fermi level separation.

The quasi-Fermi level separation can be regarded as causing the luminescent infrared photon gas. In "<u>super-equilibrium</u>", this infrared gas brightness $\mathcal{B}$, can be reinterpreted as measuring the voltage $V$ in the photovoltaic cell:

$$qV = \mu = E_g - kT \ln\left\{\frac{8\pi n^2 \nu^2}{\mathcal{B}(\nu,\mu,T)c^2} + 1\right\} \quad \ldots\ldots\ldots\ldots\ldots\ldots(2)$$

In human experience, we rarely confront light at or near thermal equilibrium. In virtually all human experience, light is much brighter than the 300K thermal equilibrium intensity, except perhaps on a very dark starless night. To the extent that there is sufficient ambient light, eq'n. (2) can assign a Chemical Potential $\mu>0$ to the light brightness we deal with in our daily lives. It is perfectly reasonable to ask on a sunny day, what is today's Chemical Potential as given by eq'n. (2)? It would be like a weather forecast, and combined with ambient temperature, would provide an upper physical limit for that day's photovoltaic cell voltage.

Luminescent infrared photons are not lost to photovoltaics. Owing to light trapping, only the small fraction $1/4n^2 \sim 2\%$ of the internal infrared luminescence escapes. 98% of the infrared luminescence is trapped upon each front surface reflection and is subject to re-absorption. Therefore modeling of solar cell performance must include a full optical analysis, on an equal footing with the analysis of minority carrier transport. Indeed, in a good solar cell, the minority carrier properties are usually close to ideal, and the performance is completely determined by the photovoltaic optics [5]. This is certainly true for the current flat-plate record solar cell, [6], which surpassed the previous record by further improving rear-surface reflectivity.

The remaining 2% of internal infrared luminescence photons, lost to escape from the front surface, --are not really lost. The front surface of a solar cell must be open to allow the entry of sunlight. At open-circuit no current is drawn. By detailed balance, [7], those incoming solar photons are replaced by outgoing luminescent photons, from the internal photon gas that is $4n^2$ times brighter than the incident sunlight. Thus the 2% photon escape is a necessity, not a loss mechanism. This also led to the slogan: "A great solar cell also needs to be a great Light Emitting Diode" [5].

In some forms of photovoltaic cell modeling, it is necessary to know the B-coefficient, where B$np$ is the rate of spontaneous emission from electron/hole recombination. But the spontaneous emission co-efficient can be bypassed by using Detailed Balance [8] to compute the spontaneous emission; where Luminescent emission≡L(ν,μ,T)=α(ν,μ,T)×$\mathcal{E}$(ν,μ,T) is exactly balanced by optical absorption of the internal luminescence photons. Then spontaneous Luminesce is:

$$L(\nu,\mu,T) = \frac{8\pi n^2 \nu^2}{c^2} \frac{\alpha(\nu,\mu,T)}{e^{(h\nu-\mu)/kT}-1} \quad\ldots\ldots\ldots\ldots\ldots\ldots\ldots\ldots\ldots(3)$$

When the material is only moderately excited, μ<<hν, the absorption spectrum is unchanged from that of unexcited material, replacing α(ν,μ,T) with α(ν,0,T). Thus the spontaneous emission can be known without a knowledge of either the B-coefficient, nor the intrinsic carrier density, $n_i$, in the semiconductor. Indeed only the conventional absorption coefficient α(ν,0,T) is needed, and eq'ns. (1-3) in this chapter apply equally well to dye molecules as to semiconductors. A dye molecule can have an internal Chemical Potential determined by the excess probability $p_{ex}$ of being in the excited state, relative to $p_o$ the usual Boltzmann probability, $\mu = kT\ ln\{p_{ex}/p_o\}$.

**4. Ergodic Light Trapping:**

Light trapping inside the solar cell improves the voltage, current, and fill-factor. But it was not exploited in the first 40 years of the photovoltaic industry. One reason is that the first generations of photovoltaic cells consisted of a plane-parallel slab. In semiconductors, the refractive index is high, $n=3.5$ in Silicon. By Snell's Law, the refraction inside the slab would always be within an angle arcsin(1/3.5)~16°, a cone adjacent to normal. As shown in Fig. 2(a), the simple geometry of a plane parallel slab fails to scatter light rays in directions outside this narrow cone which only accounts for $(1/4n^2)$~2% of $4\pi$ steradians. The other 98% of angles remain inaccessible.

This problem is solved by simply leaving the rear surface of silicon rough, as sawcut, to break the plane parallel slab symmetry. The light propagation becomes ergodic as discussed in Fig. 2(b), and shown in Fig. 1(a). Scattering the internal light by >16°, which would require an 8° rear facet, is usually sufficient. Light trapping was not generally adopted by the photovoltaic industry until the late 1990's, but it was already thoroughly exploited in Martin Green's PERC cell [9] a decade earlier. The benefit is to increase the internal optical brightness by $4n^2$, and the optical free energy by $kT\ ln\{4n^2\}$, and the photovoltaic output voltage by $(kT/q)\ ln\{4n^2\}$.

In their analysis of fundamental solar cell efficiency, Shockley & Queisser [7], S&Q, idealized the optical situation. Optical absorption jumped from zero to infinity at the bandedge. The material itself had 100% luminescence efficiency, and the luminescence was immediately re-absorbed with no losses. There was no need for a distinction between internally trapped light and Figure 2(a): Non-ergodic: The term ergodic refers to the time average trajectory of a light ray being the same as the phase-space average. The simple geometric shapes, sphere, rectangle, parallelogram fail this test. A light ray that enters such a shape but rapidly escapes, and fails to fill all possible internal angles.

Figure 2(b): Ergodic: In the case of odd shapes, the trapezoid, or the race track, light rays scatter at unusual angles, filling the full internal angle space. The light intensity builds up increasing brightness by $4n^2$, and free energy by $kT \ln\{4n^2\}$, and the photovoltaic output voltage by $(kT/q) \ln\{4n^2\}$. Almost all but the simplest geometric shapes are ergodic.

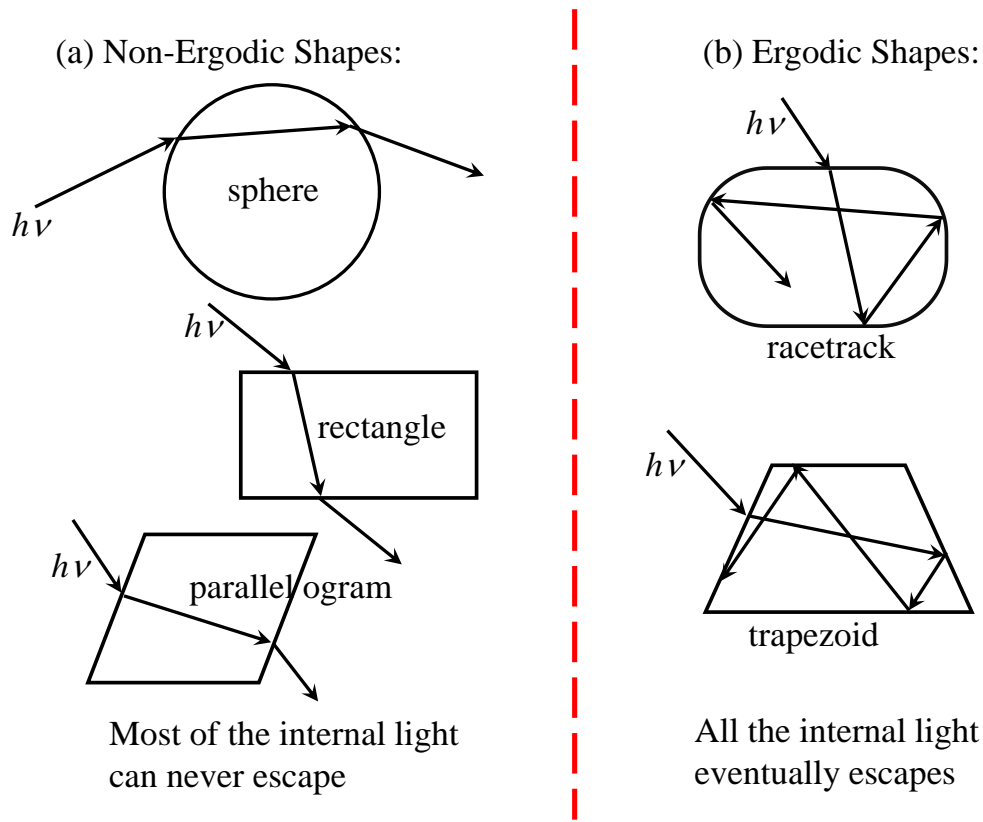

(a) Non-Ergodic Shapes:  |  (b) Ergodic Shapes:

sphere
rectangle
parallelogram

racetrack
trapezoid

Most of the internal light can never escape | All the internal light eventually escapes

external luminescence. All of the solar cell optics was effectively idealized away, and there was no need to talk about the two internal photon gases.

S&Q achieved their goal of placing an upper limit on solar efficiency. The purpose of this article is to consider the parasitic optical effects, and to show that these deleterious effects can be largely overcome by proper optical management and design. By maximizing the optical reflectivity in the context of ergodic optics, and selecting materials with very high internal luminescence yield, we can approach the idealized S&Q performance.

**5. Non-ideal Solar Cells:**

We now consider a series of non-ideal solar cells Figs. 3(b-d). But Fig. 3(a), is an exception. In 3(a) we have done the best possible job with the optics. There is a high reflectivity mirror, so that none of the photons are lost. Ergodicity is provided in two ways: by the textured rear reflector, and by the re-absorption and random angle re-emission of the infrared luminescence. The spectral shape of optical absorption and re-emission can be taken into account using the Shockley-Van Roosbroeck [8] principle. The result for GaAs is 33.5% efficiency [5] for standard Air Mass 1.5 illumination [10], similar to the S&Q result, but without the idealistic assumptions. In ref. [5] we benefit from the very low Auger recombination relative to internal luminescence in direct-gap GaAs.

The non-idealities in GaAs solar cells prior to 2011 are represented in Fig. 3(b). The material was of top epitaxial quality, but the films rested on the original growth substrates. The substrate was thick, relatively impure, and had no proper rear mirror. The optical quality was poor. The substrate was essentially a sink for bandedge luminescence. This immediately wasted 98% of the luminescent photons, like losing 98% of the minority carriers at open-circuit. The main effect was to lose $\sim(kT/q) \, ln\{4n^2\}$ in *Voc*, open-circuit voltage. Eliminating this problem by epitaxial liftoff was the reason for all the new solar cell efficiency records in the 2010's decade.

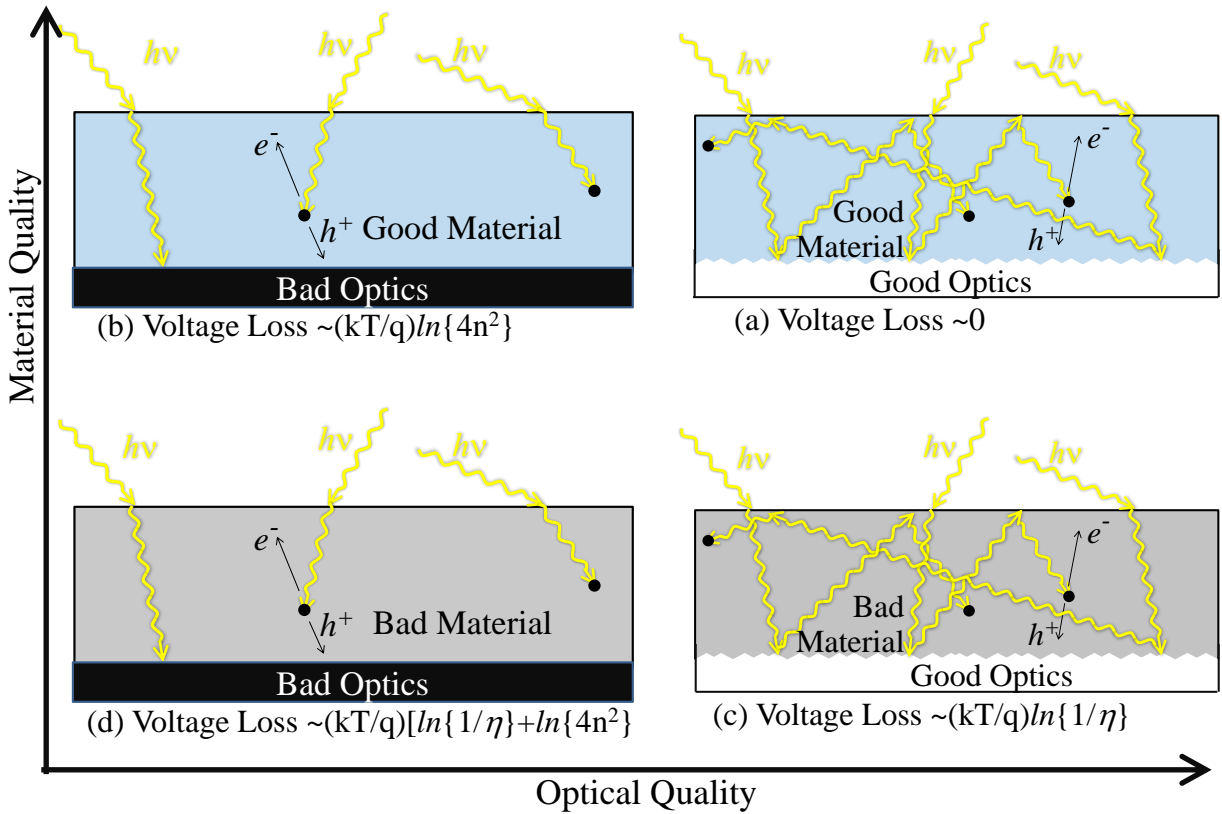

Figure 3: A comparison of the effect of Light Trapping Optics on the operating point voltage penalty $\Delta V_{OP}$ from poor internal luminescence efficiency, $\eta$, and/or poor optical design.

Case 3(a) is the ideal case of a perfect scattering rear mirror for ergodicity, combined with the record-breaking internal luminescence efficiency of epitaxial thin GaAs film separated from their growth substrates by epitaxial liftoff.

Case 3(b) represents epitaxial GaAs solar cells before 2011, which were still attached to the original GaAs growth substrates, that were so thick as to effectively absorb bandedge luminescence.

Case 3(c) is a modern Si solar with good light management, but the material is labelled bad since the indirect gap prevents high (>90%) internal luminescence efficiency.

Case 3(d) represents the older generation of Silicon solar cells, (pre-1990), still indirect, but also with no implementation of light trapping.

Consider the opposite non-ideality in Fig. 3(c). The optics are good as in modern c-Si solar cells, but the material is handicapped by being indirect. Although Si is indirect, it still luminesces, and can be an LED [11]. Nonetheless, Si is not efficient enough to build up a high brightness of luminescent photons, which requires an internal luminescence efficiency >90% owing to the multiple absorption/re-emission events. The loss of luminescent photons reduces luminescent brightness and by eq'n. (2) leads to a corresponding drop in open-circuit voltage $V_{oc}$ by $\sim(kT/q)\ ln\{1/\eta\}$, where $\eta$ is the internal luminescence yield in the material.

But there is ample motivation for good optical design, in Fig. 3(c) despite Si being an indirect semiconductor. Good optical design increases the effective optical absorption coefficient by $\sim 4n^2 \sim 50$, and allows the Silicon to be thinner by the same factor. The photo-carriers from the sun are effectively compressed into a thinner layer, and reside at a higher density by $\sim 4n^2$ times. The higher density corresponds to less entropy and more voltage. The designer who uses good optics to make his Si-solar cell thinner would then gains $\sim(kT/q)\ ln\{4n^2\}$ in $V_{oc}$. But good optical design is already assumed in both Figs. 3(a) & 3(c).

Such good optical design is one of the reasons that Si-solar cells are far exceeding the theoretical limits that were projected in 1979 [12, 13]. The other reason is that higher performing cells require good material quality. But good material quality also includes low recombination [14] on the surfaces and interfaces. This was first achieved by Swanson [15] using point contact openings on oxidized Si, effectively creating the first double heterostructure [16,17] on Silicon. The oxide coating was particularly compatible with good optical mirror reflectivity.

The remaining case is Fig. 3(d), corresponding to the old-fashioned Silicon solar cells before light-trapping and Si heterojunctions became standard. In those days, the rear reflectivity was rather poor, i.e. bad optics, and the material being indirect was yet a second penalty. The material had to be thick enough to absorb the solar photons, with no help from light trapping. A thicker solar cell means a lower minority carrier concentration, since the same injected carriers occupy a larger volume. Thus the open-circuit voltage is penalized by $-(kT/q) [ln\{1/\eta\} + ln\{4n^2\}]$, once for the poor luminescence efficiency, and again for poor optical absorption, necessitating greater thickness.

## 6. Conclusions:

For the reasons given in this paper, the photovoltaic industry employs solar light-trapping in almost all cases. But there is now a second mechanism with the recognition of the internal infrared luminescent photon gas that can build up to a high brightness, many suns, inside direct bandgap materials. If this infrared luminescent photon gas is properly conserved, absorbed, re-emitted, and reflected, then the Shockley-Queisser limit can be approached even under a realistic practical optical design.

## 7. Acknowledgement:

We have benefited from many insights and suggestions by the solar cell authority, Dick Swanson.

This research was supported by Department of Energy (DOE) "Light-Material Interactions in Energy Conversion" Energy Frontier Research Center under Grant DE-SC0001293, DOE "Photonics at Thermodynamic Limit" Energy Frontier Research Center under Grant DE-SC00019140.

14. If the surfaces and interfaces dominate the electron-hole recombination, there would be no benefit in making the solar cell thinner.